\documentclass[twocolumn,prl,superscriptaddress,groupedaddress]{revtex4}  
\usepackage{graphicx}  
\usepackage{amssymb}   
\usepackage{amsmath}

\newcommand*\xbar[1]{%
  \hbox{%
    \vbox{%
      \hrule height 0.5pt 
      \kern0.5ex
      \hbox{%
        \kern-0.1em
        \ensuremath{#1}%
        \kern-0.1em
      }%
    }%
  }%
} 

\begin{document}

\title{Accurate sensing of multiple ligands with a single receptor}

\author{Vijay Singh} \affiliation{Department of Physics, Emory
  University, Atlanta, GA 30322, USA}
\author{Ilya  Nemenman} \affiliation{Department of Physics, Emory
  University, Atlanta, GA 30322, USA}
\affiliation{Department of Biology, Emory
  University, Atlanta, GA 30322, USA}

\begin{abstract} 
  Cells use surface receptors to estimate the concentration of
  external ligands. Limits on the accuracy of such estimations have
  been well studied for pairs of ligand and receptor species.
  However, the environment typically contains many
  ligands, which can bind to the same receptors with different
  affinities, resulting in cross-talk. In traditional rate models, such cross-talk prevents accurate
  inference of individual ligand concentrations. In contrast, here we
  show that knowing the precise timing
  sequence of stochastic binding and unbinding events allows one
  receptor to provide information about
  multiple ligands simultaneously and with a high accuracy. We argue
  that such
  high-accuracy estimation of multiple concentrations can be realized
  by the familiar kinetic proofreading mechanism.
\end{abstract}

\maketitle

{\em Introduction}: Cells obtain information about their environment
by capturing ligand molecules with receptors on their surface and
estimating the ligand concentration from the receptor activity. Limits
on the accuracy of such estimation have been a subject of interest
since the seminal work of Berg and Purcell \cite{Berg:1977bp}, with
several substantial extensions found recently
\cite{bray1998receptor,Bialek2005,Endres:2008eb,endres2009maximum,hu2010physical,kaizu2014berg,mugler15}.
All of these assume one ligand species coupled to one receptor
species. However, cells carry many types of receptors and have many
species of ligands around them. The same ligands can bind to many
receptors, albeit with different affinities, and vice versa. This is
commonly referred to as {\em cross-talk}.

In traditional deterministic chemical kinetics, one cannot estimate
concentrations of more ligands than there are receptor types. Further,
even a weak cross-talk prevents determination of concentrations of
individual chemical species since activity of a receptor is a function
of a weighted sum of concentrations of all ligands that can bind to
it. In contrast, here we argue that, with cross-talk, concentration of
more than one chemical species can be inferred from the activity of
one receptor, provided that the entire stochastic temporal sequence of
receptor binding and unbinding events is accessible instead of its
mean occupancy. This surprising result can be understood by noting
that a typical duration of time that a ligand remains bound to the
receptors depends on its unbinding rate. Thus observing the statistics
of the receptor's unbound time durations allows estimation of a
weighted average of all chemical species that interact with it
\cite{endres2009maximum}, and then observing the statistics of the
bound time durations allows to tell how common each ligand is.

In this article, we derive these results for the simplest problem of
the class, namely one receptor interacting with two ligand species.
While the exact solution of the inference problem for finding both
ligand concentrations is hard to implement using common biochemical
machinery, we show that an accurate approximation is possible using
the familiar kinetic proofreading mechanism
\cite{hopfield1974kinetic,ninio1975kinetic}.

\begin{figure}[t] \begin{center} \includegraphics[scale=0.32,trim=
    0.4cm 0.0cm 0.0mm 0.0cm]{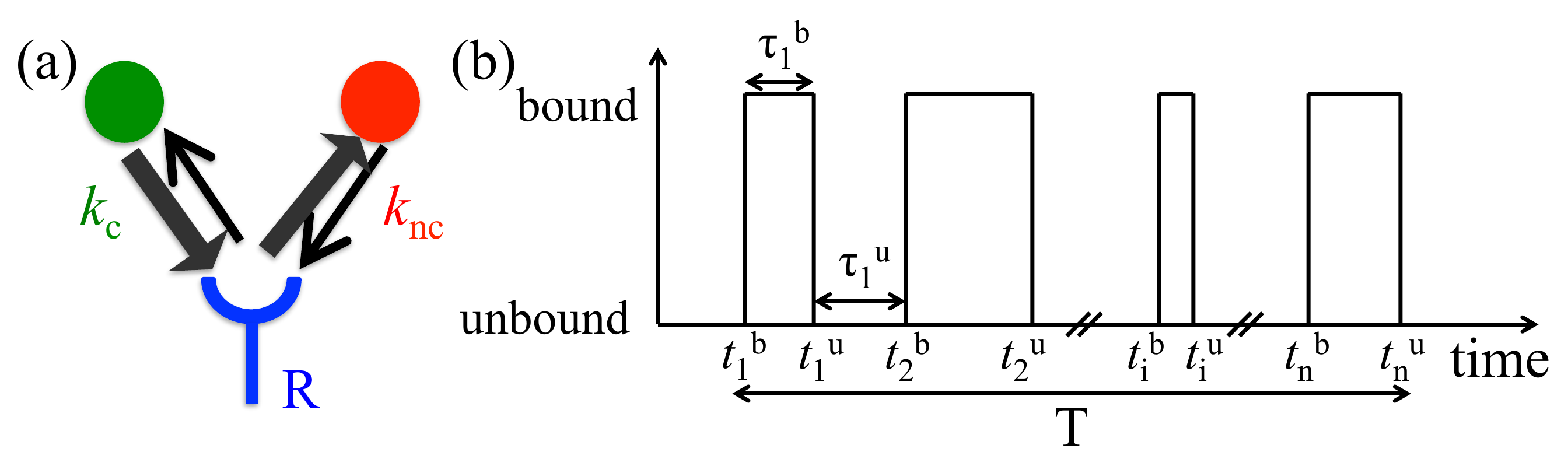} \caption{{\bf The model.}
      (a). Two lignads, cognate and non-cognate, bind to a receptor R
      with binding rates $k_{\rm{c}}$ and $k_{\rm{nc}}$,
      respectively. The cognate unbinding rate is defined as lower
      than the non-cognate one ($r_{\rm{c}} < r_{\rm{nc}}$). (b) Time
      series of receptor occupancy is used to determine both
      on-rates.} \label{model}
  \end{center} \end{figure} {\em The Model}: Consider a single
receptor estimating concentrations of a cognate and a non-cognate
ligand, Fig.~\ref{model}. The ligands bind to the receptor with
on-rates $k_{\rm{c}}$ and $k_{\rm{nc}}$. These are proportional to the
ligand concentrations with known coefficients of proportionality. Thus
estimating $k_{\rm c,nc}$ is equivalent to estimating the
concentrations themselves. The unbinding, or off-rates, $r_{\rm{c}}$
and $r_{\rm{nc}}$, distinguish the two ligands:
$r_{\rm nc}>r_{\rm c}$, and a cognate molecule typically stays bound
for longer. Following Ref.~\cite{endres2009maximum}, we estimate
$k_{\rm c}$ and $k_{\rm nc}$ from the time-series of binding,
$\{t^{\rm b}_i\}$, and unbinding, $\{t^{\rm u}_i\}$ events of a total
duration $T$ using Maximum Likelihood techniques. The numbers of
binding and unbinding events are different by, at most, one, which is
insignificant since we consider $T\to\infty$. Thus without loss of
generality, we assume that the first event was a binding event at
$t^{\rm b}_1$, and the last one was the unbinding at $t^{\rm u}_n$. We
write the probability distribution of observing the sequence
$\{t^{\rm b}_1,t^{\rm u}_1,\dots,t^{\rm b}_n,t^{\rm u}_n\}$, or
alternatively the sequence of binding and unbinding intervals
$\tau^{\rm b}_i=t^{\rm u}_i-t^{\rm b}_i$, and
$\tau^{\rm u}_i=t^{\rm b}_{i+1}-t^{\rm u}_i$:
 \begin{multline}
P\equiv P(\{\tau^{\rm b}_i,\;\tau^{\rm
   u}_i\}|k_{\rm{c}},k_{\rm{nc}})=\frac{1}{Z}\prod_{i=1}^{n} \left[
                                   e^{-\tau^{\rm u}_{i} (k_{\rm{c}} +  k_{\rm{nc}})} \right.\\
        \left.\times \left( k_{\rm{c}} \; r_{\rm{c}} \; e^{-\tau^{\rm b}_i
          r_{\rm{c}}} +k_{\rm{nc}}\; r_{\rm{nc}} \; e^{-\tau^{\rm b}_i
          r_{\rm{nc}}} \right) \right].
\label{eq:prob}
\end{multline}
Here the first term under the product sign is the probability of the
receptor staying unbound for $\tau^{\rm u}_i$. The second term, which
we from now on denote by $D(k_{\rm c},k_{\rm nc},\tau^{\rm b}_i)$, is
proportional to the probability of staying bound for $\tau^{\rm b}_i$,
which has contributions from being bound to the cognate and the
noncognate ligands, with odds of $k_{\rm c}/k_{\rm nc}$.  Finally, $Z$
is the normalization. Note that here we define
$\tau^{\rm u}_n=t^{\rm b}_1+(T-t^{\rm u}_n)$, so that the $n$'th
unbound interval includes the ``incomplete'' unbound intervals before
the first binding and after the last unbinding.

The log-likelihood of $k_{\rm c,nc}$ is the logarithm of $P$,
Eq.~(\ref{eq:prob}). Taking the derivatives of the log-likelihood
w.~r.~t.~$k_{\rm c}$ and $k_{\rm nc}$ and setting them to zero gives
the Maximum Likelihood (ML) equations for the two concentrations.
Denoting by $T^{\rm u}=\sum_{i=1}^n\tau^{\rm u}_i$ the total time the
receptor is unbound, these are 
\begin{align}
  -T^{\rm u} + \sum_{i=1}^{n} \frac{r_{\rm{c}} e^{-\tau^{\rm b}_i
  r_{\rm{c}}}} {D(k^*_{\rm c},k^*_{\rm nc},\tau^{\rm b}_i) } &=0, \label{deq1} \\ 
 -T^{\rm u} + \sum_{i=1}^{n} \frac{ r_{\rm{nc}} e^{-\tau^{\rm b}_i
  r_{\rm{nc}}}} {D(k^*_{\rm c},k^*_{\rm nc},\tau^{\rm b}_i)} &=0 \label{deq2},
\end{align} 
where $*$ denotes the ML solution. Multiplying Eqs.~(\ref{deq1},
\ref{deq2}) by $k_{\rm c}^*$ and $k_{\rm nc}^*$, respectively, and
adding them gives
\begin{equation}
  k_{\rm c}^*+k_{\rm nc}^* =\frac{n}{T^{\rm u}},
\label{eq:Tu}
\end{equation} 
which determines the sum of the two concentrations, showing that the
estimates are negatively correlated. As in
Ref.~\cite{endres2009maximum}, the total on-rate (the weighted average
of the external concentrations) is determined only by the average
duration of the unbound interval, $(n/T^{\rm u})^{-1}$, because no
binding is possible when the receptor is already bound.

In general, the ML equations cannot be solved analytically, requiring
numerical approaches. However, as all ML estimators, they are unbiased
to the leading order in $n$. The standard errors of the ML estimates can
be obtained by inverting the Hessian matrix,
 \begin{multline}
 \left. \frac{\partial^2 \log P}{\partial
     k_{\cdot}\partial{k_{\cdot}}}\right|_{k^*_{\rm c},k^*_{\rm nc}}=
  \sum_{i=1}^n\left[\frac{-1}{D(k_{\rm c},k_{\rm nc},\tau^{\rm b}_i)^2}\right.\\
\left.\times\left(\begin{array}{ccc}r^2_{\rm c}e^{-2\tau^{\rm
                    b}_ir_{\rm c}}& r_{\rm c}r_{\rm nc}e^{-\tau^{\rm
                    b}_i(r_{\rm c}+r_{\rm nc})}\\
r_{\rm c}r_{\rm nc}e^{-\tau^{\rm
                    b}_i(r_{\rm c}+r_{\rm nc})} &r^2_{\rm nc}e^{-2\tau^{\rm
                    b}_ir_{\rm nc}}
\end{array}\right) \right],
\label{eq:hessian}
\end{multline}
where $\cdot$ stands for $\{{\rm c},{\rm nc}\}$. The inverse of
$\frac{\partial^2 \log P}{\partial k_{\cdot}\partial{k_{\cdot}}}$,
which scales as $\propto 1/n$, sets the minimum variance of any
unbiased estimator according to the Cramer-Rao bound. It has
straightforward analytical approximations in various regimes. For
example, for $k_{\rm c}/k_{\rm nc}\gg1$ and
$r_{\rm c}/r_{\rm nc}\ll1$, when the noncognate ligand is almost
absent, and its few molecules do not bind for long, one gets
$\sigma^2(k^*_{\rm c})\approx \left(\partial^2\log P/\partial k_{\rm
    c}^{2}\right)^{-1}_{k_{\rm c}=k_{\rm c}^*}\approx1/n$,
matching the accuracy of sensing one ligand with one receptor
\cite{endres2009maximum}. A regime relevant for detection of a rare,
but highly specific ligand \cite{Lalanne:2015ut,Francois:2013gx}) can
be investigated as well.  Instead, we focus on how the receptor
estimates (rather than detects) concentrations of {\em both} ligands
simultaneously, which requires us to investigate the full range of
on-rates.

\begin{figure*}
\begin{center} \includegraphics[scale=0.5]{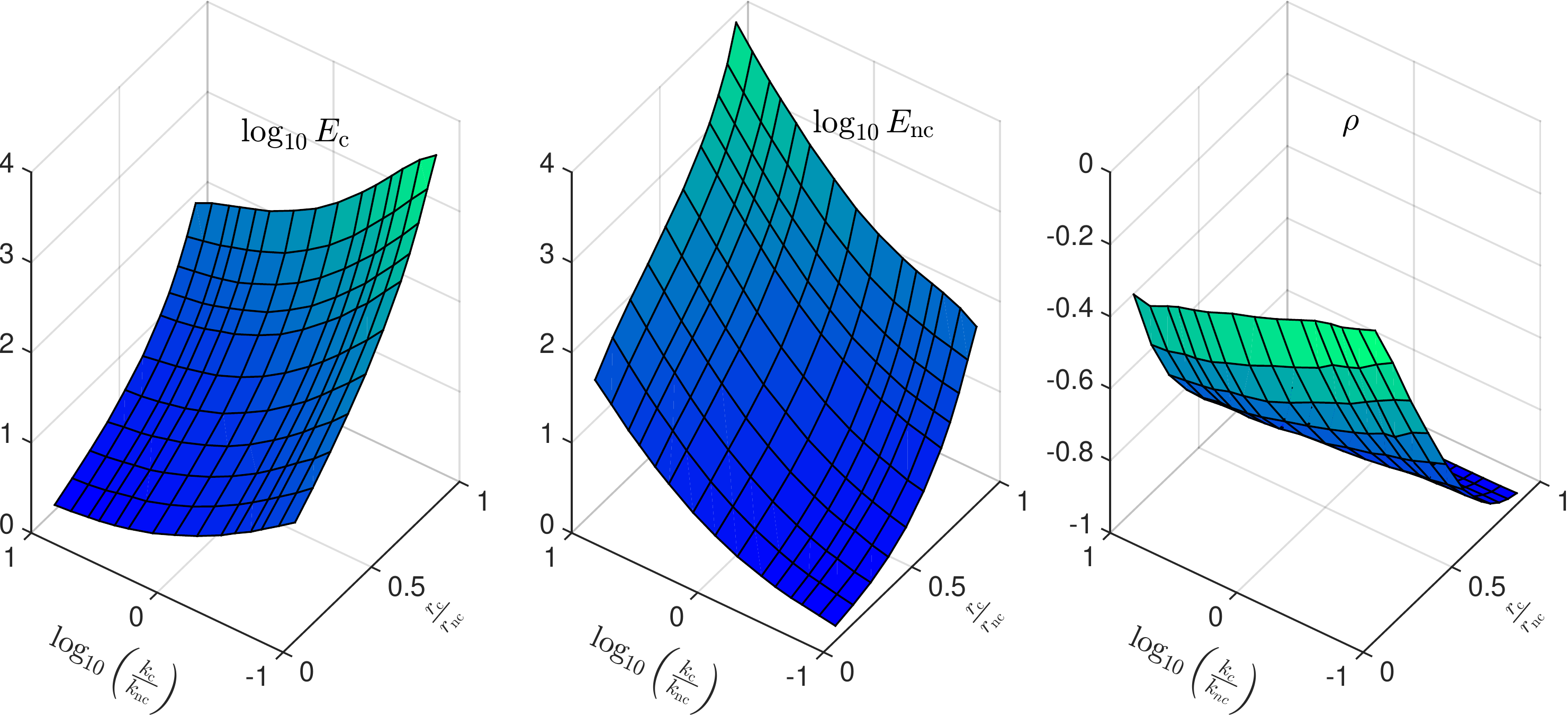}
  \caption{{\bf Variability of the ML estimators}, represented by
    $\log_{10} E_{\rm c}$ (left), $\log_{10}E_{\rm nc}$ (center), and
    the correlation coefficient $\rho$ between $k_{\rm c}^*$ and
    $k_{\rm nc}^*$ (right) as functions of $k_\cdot$ and
    $r_\cdot$. Here we use $r_{\rm nc}=k_{\rm c}+k_{\rm nc}=1$. The
    plotted quantities are estimated as averages over $30,000$
    randomly generated binding/unbinding sequences for each
    combination of the rates. Each sequence consists of $n=30,000$
    binding events, simulated using the Gillespie algorithm. Standard
    errors are too small to be represented.}
\label{fig:ML}
\end{center}
\end{figure*}

To study the variability of the ML estimator, we define its error as
$E_{\rm c,nc}=n\sigma^2(k_{\rm c,nc}^*)/k_{\rm c,nc}^2$, the squared
coefficient of variation, multiplied by $n$, which has a finite limit
at $n\to\infty$.  $E=1$ corresponds to the accuracy that a receptor
measuring a single ligand would obtain \cite{endres2009maximum}. We
show $\log_{10}E$ for different on- and off-rates in
Fig.~(\ref{fig:ML}). If the two ligands are readily distinguishable,
$r_{\rm c}\ll r_{\rm nc}$, then the ligand with the dominant $k$ has
$E\sim 1$.  When $k_{\rm c}\sim k_{\rm nc}$, $E_\cdot\sim 4\dots 5$,
and it grows to $10\dots30$ for a ligand with a very small relative
on-rate. Emphasizing the importance of the time scale separation,
$E> 100$ if the ligands are hard to distinguish,
$r_{\rm c} \sim r_{\rm nc}$. Here, in addition, the correlation
coefficient $\rho$ of the two estimates reaches $-1$ because the same
binding event can be attributed to either ligand. Finally, the
asymmetry of the plots w.~r.~t.\ the exchange of $k_{\rm c}$ and
$k_{\rm nc}$ is because the cognate ligand can generate short binding
events, while long events from the noncognate ligand are exponentially
unlikely.  In summary, it is possible to infer two ligand
concentrations from one receptor, with the error of only $1\dots10$
times larger than for ligand-receptor pairs with no cross talk, as
long as the two off-rates are substantially different.

{\em Approximate solution}. Solving Eqs.~(\ref{deq1},~\ref{deq2}) to
find the ML on-rates would be hard for the cell. Luckily, an
approximate solution exists. To find it, we notice that most of the
long binding events come from the cognate ligand since the noncognate
one dissociates faster. Defining long events as
$\tau^{\rm b}_i\ge T^{\rm c}$, we rewrite Eqs.~(\ref{deq1},
\ref{eq:Tu}) as \begin{equation} \frac{n}{k^*_{\rm{c}}+k_{\rm{nc}}^*}
  = \left(\sum_{\tau^{\rm b}_i\ge T^{\rm c}} +\sum_{\tau^{\rm b}_i<
      T^{\rm c}} \right)\frac{ r_{\rm{c}} e^{-\tau^{\rm b}_i
      r_{\rm{c}}}} {D(k^*_{\rm c},k^*_{\rm nc},\tau^{\rm b}_i)}
\end{equation} 
Assuming that almost all long events are cognate,
$T^{\rm c}\gg 1/r_{\rm nc}$, this gives
\begin{equation}
  \frac{n}{k^{\rm a}_{\rm{c}}+k_{\rm{nc}}^{\rm a}} =\frac{n_{\rm
      l}}{k^{\rm a}_{\rm c}} +\sum_{\tau^{\rm b}_i< T^{\rm c}} \frac{
    r_{\rm{c}} e^{-\tau^{\rm b}_i r_{\rm{c}}}} { D(k^{\rm a}_{\rm
      c},k^{\rm a}_{\rm nc},\tau^{\rm b}_i) },
\label{approx:sum}
\end{equation}
where $n_{\rm l}$ is the number of long events, and the superscript
``a'' stands for the {\em a}pproximate solution. If further $T$ is
long enough so that there are many short events, and a single binding
duration hardly affects $k^*_{\rm c}$, then the sum in
Eq.~(\ref{approx:sum}) can be approximated by the expectation value:
 \begin{multline}
   \frac{n}{k^{\rm a}_{\rm{c}}+k_{\rm{nc}}^{\rm a}} =\frac{n_{\rm
       l}}{k^{\rm a}_{\rm c}}+ (n-n_l) \int_{0}^{T_c} \frac{
     r_{\rm{c}} e^{-\tau^{\rm b} r_{\rm{c}}}P(\tau^{\rm b}|k^{\rm
       a}_{\rm c},k^{\rm a}_{\rm nc})d \tau^{\rm b}} {D(k^{\rm a}_{\rm
       c},k^{\rm a}_{\rm nc},\tau^{\rm b}) },
\label{sum2int}
\end{multline} 
where $P(\tau^{\rm b}|k^{\rm a}_{\rm c},k^{\rm a}_{\rm nc})$ is the probability of observing a binding
event of the duration $\tau^{\rm b}$ for the given binding rates,
\begin{align} P(\tau^{\rm b}|k^{\rm a}_{\rm c},k^{\rm a}_{\rm
  nc})=\frac{ D(k^{\rm a}_{\rm
       c},k^{\rm a}_{\rm nc},\tau^{\rm b}) }{k^{\rm a}_{\rm{c}} + k^{\rm a}_{\rm{nc}}}.
  \label{prob_tb} \end{align} 
Plugging Eq.~(\ref{prob_tb}) into Eq.~(\ref{sum2int}), we obtain
\begin{align} \frac{1}{k^{\rm a}_{\rm{c}}+k^{\rm a}_{\rm{nc}}} =
  \frac{n_{\rm l}}{n k^{\rm a}_{\rm{c}}}+ \left( 1-\frac{n_{\rm l}}{n} \right)
  \frac{1-e^{-r_{\rm{c}}T^{\rm c}}} {k^{\rm a}_{\rm{c}}+k^{\rm a}_{\rm{nc}}}.
\end{align} 
Finally, since $n_{\rm l}\ll n$, using Eq.~(\ref{eq:Tu}), we get:
\begin{align} k^{\rm a}_{\rm{c}} &=\frac{n_{\rm l}}{T^{\rm u}}
                                   e^{r_{\rm{c}}\;T^{\rm c}},\label{kca}\\
  k^{\rm a}_{\rm nc}&= \frac{n}{T^{\rm u}}-\frac{n_{\rm l}}{T^{\rm u}}
                      e^{r_{\rm{c}}\;T^{\rm c}}\label{knca}.
\end{align}
In other words, the approximate cognate ligand concentration is
proportional to the number of long events.

We can estimate the bias and the variance of $k_{\rm c,nc}^{\rm a}$ in
a limiting case. If $r_{\rm c}$ and $r_{\rm nc}$ are not very
different from each other, then $T^{\rm c}$ must be much larger than
the inverse of either of them,
$T^{\rm c}\gg\{r^{-1}_{\rm nc},r_{\rm c}^{-1}\}$, and
$n_{\rm l}\ll n$. Then most of the variance of $k_{\rm c,nc}^{\rm a}$
in Eqs.~(\ref{kca}, \ref{knca}) comes from variability of $n_{\rm l}$,
but not $T^{\rm u}$. Thus we write
$\langle k_{\rm c}^{\rm a}\rangle \approx \frac{\langle n_{\rm
    l}\rangle}{\langle T^{\rm u}\rangle} e^{r_{\rm c}T^{\rm c}}$.
Further, the individual unbound periods are independent, so that
$\langle T^{\rm u}\rangle=n\langle \tau^{\rm u}\rangle=n/(k_{\rm
  c}+k_{\rm nc})$
(notice the use of $k$ rather than $k^{\rm a}$ in this
relation). Further,
$\langle n_{\rm l}\rangle=n\, P(\tau^{\rm b}>T^{\rm c})=
\frac{n}{k_{\rm c}+k_{\rm nc}}\left(k_{\rm c}e^{-r_{\rm c}T^{\rm c}}+
  k_{\rm nc}e^{-r_{\rm nc}T^{\rm c}}\right)$.
Combining these expressions, we get
\begin{equation}
\langle k_{\rm c}^{\rm a}\rangle\approx k_{\rm c}+k_{\rm nc}e^{-(r_{\rm
    nc}-r_{\rm c})T^{\rm c}}.
\label{bias}
\end{equation}
Thus for large $T^{\rm c}$, the bias of the approximate estimator,
$k_{\rm nc}e^{-(r_{\rm nc}-r_{\rm c})T^{\rm c}}$, grows with the
relative number of noncognate long bindings events. In turn, the
latter is proportional to $k_{\rm nc}$, but decreases exponentially
with $T^{\rm c}$.

Within the same approximation, the variance of the estimator is
$\sigma^2( k_{\rm c}^{\rm a}) \approx \frac{\sigma^2(n_{\rm
    l})}{\langle T^{\rm u}\rangle^2} e^{2r_{\rm c}T^{\rm c}}$.
But long binding events are rare, independent of each other, and hence
obey the Poisson statistics. Thus
$\sigma^2(n_{\rm l})=\langle n_{\rm l}\rangle$, so that
\begin{equation}
  \sigma^2( k_{\rm c}^{\rm a}) \approx \langle k_{\rm c}^{\rm a}\rangle \,\frac{k_{\rm c}+k_{\rm
    nc}}{n}e^{r_{\rm c}T^{\rm c}}.
\label{eq:var_app}
\end{equation}
The variance obviously grows with $T^{\rm c}$.

Knowing that the bias and the variance of the approximation change in
opposite directions with $T^{\rm c}$, we can find the optimal cutoff
by minimizing the overall error, or, in other words, solving the
bias-variance tradeoff:
\begin{equation}
  T^{\rm c}_*=\arg \min_{T^{\rm c}} L=\arg \min_{T^{\rm c}}  \left[\left(k_{\rm c}-\langle k_{\rm c}^{\rm a}\rangle\right)^2+\sigma^2(k^{\rm
      a}_{\rm c})\right],
  \label{tradeoff}
\end{equation}
where $L$ is the sum of the squared bias and the variance of the
estimator.  Near the optimal cutoff, the bias is small, and we use
$k_{\rm c}$ instead of $k_{\rm c}^{\rm a}$ for the variance of the
estimator, Eq.~(\ref{eq:var_app}). Then solving Eq.~(\ref{tradeoff})
gives:
\begin{equation}
T^{\rm c}_*= \frac{1}{(2 r_{\rm nc}- r_{\rm c})}\log\left[ 2 T^{\rm u} \left(\frac{r_{\rm nc}}{r_{\rm c}}-1 \right) \frac{k_{\rm nc}^2}{k_{\rm c}} \right].
\end{equation}
Plugging this into Eqs.~(\ref{bias},~\ref{eq:var_app}), we can get the
minimal error of the estimator, which we omit here for brevity.

\begin{figure*}
\begin{center} \includegraphics[scale=0.5]{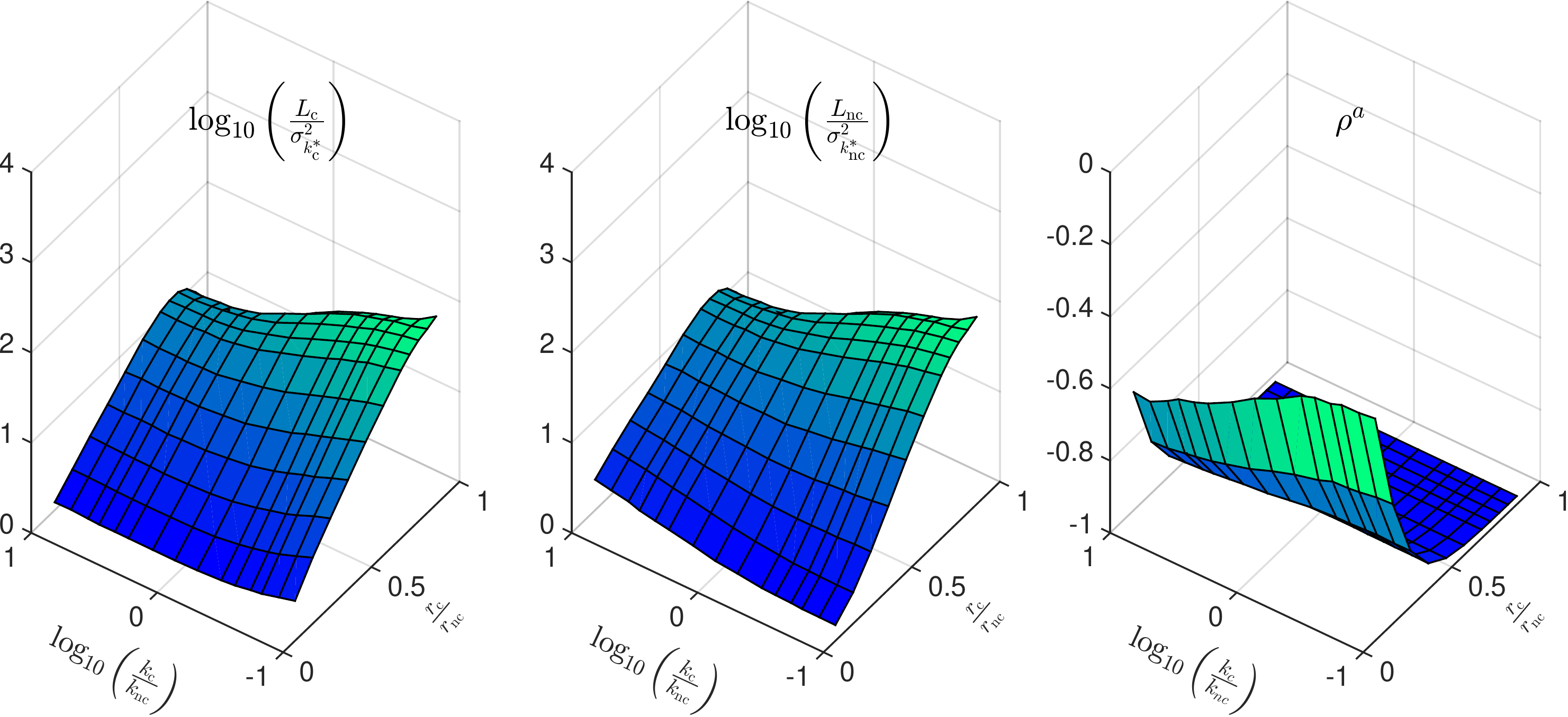}
  \caption{{\bf Comparison of errors of the approximate and the ML
      solutions}. We plot
    $\log_{10}({L_{\rm c}(T_0)}/{\sigma^2_{\rm k^*_{\rm c}}})$ (left),
    $\log_{10}({L_{\rm nc}(T_0)}/{\sigma^2_{\rm k^*_{\rm nc}}})$
    (center) and the covariance of the approximate estimates (right)
    as functions of on- and off-rates. Simulations are performed in
    the same way as in Fig.~\ref{fig:ML}. }
\label{fig:Loss}
\end{center}
\end{figure*}

The optimal cutoff is $\propto1/r_{\rm nc}$ if
$r_{\rm nc}\gg r_{\rm c}$, and it grows with $r_{\rm c}$, allowing for
better disambiguation of cognate and noncognate events.  Crucially,
the off-rates are specified with the ligand identities. In contrast,
the on-rates, $k_{\rm c,nc}$, are what the receptors
measures. Therefore, it is encouraging that $T^{\rm c}$ depends only
logarithmically on the on-rates (and also on the duration of the
measurement, $T^{\rm u}$): fixing $T^{\rm c}$ as $T^{\rm c}_*$ at some
fixed values of $k_{\rm c,nc}$ remains near-optimal for a broad range
of on-rates. To illustrate this, we use
$T^{\rm c}=T^{\rm c}_*(k_{\rm c}=k_{\rm nc}=1/2)\equiv T_0$ and
analyze the quality of the approximation in Fig.~\ref{fig:Loss}, where
we plot the ratio $L_{\rm c,nc}(T_0)/\sigma^2_{k_{\rm c,nc}}$.  Since
the ratio approaches 1 when $r_{\rm c}/ r_{\rm nc}\to 0$
(specifically, for $r_{\rm c}/r_{\rm nc}=0.1$,
$L_{\rm c}(T_0)/\sigma^2_{k_{\rm c}}\approx 1.47$, and
$L_{\rm nc}(T_0)/\sigma^2_{k_{\rm nc}}\approx 1.21$), we conclude that
the approximation is accurate even at fixed $T^{\rm c}=T_0$ when its
assumptions are satisfied. In contrast, when the ligands are nearly
indistinguishable,
$L_{\rm c,nc}(T_0)/\sigma^2_{k_{\rm c,nc}}\sim 100$, but here one
would not use one receptor to estimate two concentrations since even
the ML solution is bad (cf.~Fig.~\ref{fig:ML}). Note also that both
$L_{\rm c}$ and $L_{\rm nc}$ are smaller for
$r_{\rm c}\sim r_{\rm nc}$ if $k_{\rm c}\gg k_{\rm nc}$. This is
because our main assumption (that almost all long events are cognate)
holds better when cognate ligands dominate. Finally, the correlation
coefficient between the approximate estimates, $\rho^{\rm a}$ (right
panel) reaches -1 earlier than in Fig.~\ref{fig:ML}. This is a direct
consequence of Eqs.~(\ref{kca}, \ref{knca}).

{\em Kinetic Proofreading for approximate estimation}. The approximate
solution can be computed by cells using the well-known kinetic
proofreading (KPR) mechanism
\cite{hopfield1974kinetic,ninio1975kinetic,Mckeithan95,Goldstein08}.
In the simplest model of KPR \cite{Bel10}, intermediate states between
an inactive and an active state of a receptor delay the
activation. Thus bound ligands can dissociate before the receptor
activates, at which point it quickly reverts to the inactive
state. Since $r_{\rm c}>r_{\rm nc}$, cognate ligands dominate among
bindings that actually lead to activation. The resulting increase in
specificity in various KPR schemes has led to their exploration in the
context of {\em detection} of rare ligands
\cite{Goldstein08,Francois:2013gx,Lalanne:2015ut}, and here we extend
them to {\em measurement} of concentration of cognate and noncognate
ligands simultaneously.

\begin{figure}
\begin{center} \includegraphics[scale=0.4]{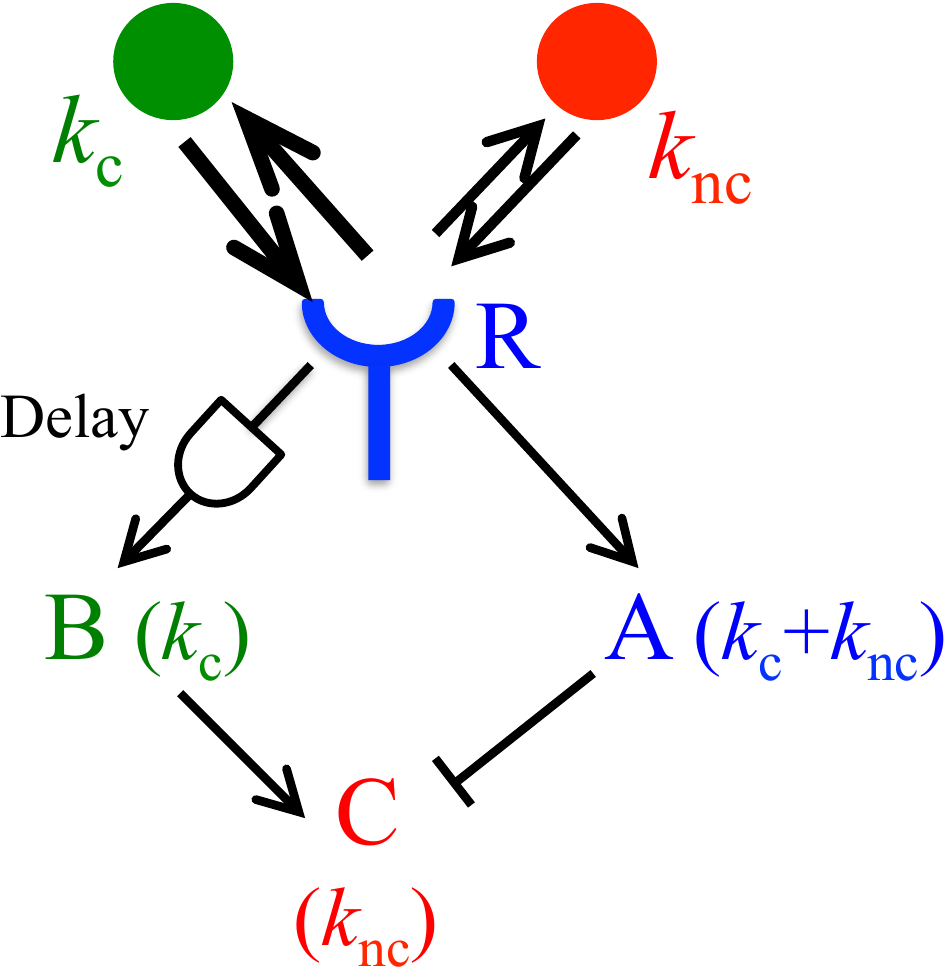}
  \caption{{\bf Kinetic Proofreading for estimating multiple
      concentrations.} Molecules A and B are produced when the
    receptor is bound, but A is produced only for long
    bindings. Another chemical species C subtracts A from B, so that A
    approximates $k_{\rm c}$ and C approxiates $k_{\rm nc}$.}
\label{fig:KPR}
\end{center}
\end{figure}

Consider a biochemical network in Fig.~\ref{fig:KPR}: the receptor (R)
activates two messenger molecules (A) and (B). The first one is
activated with the rate $k_{\rm A}$ whenever the receptor is
bound. The second one is activated only if the receptor stays bound
for longer than a certain $T^{\rm c}$ (with the delay achieved using
the KPR intermediate states). The activation rate after the delay is
$k_{\rm B}$. The molecules deactivate with the rates $r_{\rm A}$ and
$r_{\rm B}$, respectively, and all activations/deactivations are
first-order reactions. Then the mean concentrations of the messenger
molecules are:
\begin{align}
\bar{A}&=\frac{k_{\rm c}/r_{\rm c}+k_{\rm nc}/r_{\rm nc}}{1+k_{\rm
  c}/r_{\rm c}+k_{\rm nc}/r_{\rm nc}}\frac{k_{\rm A}}{r_{\rm A}}\label{eq:A}\\
\bar{B}&=\frac{k_{\rm c}/r_{\rm c}e^{-r_{\rm c}T^{\rm c}}+k_{\rm
         nc}/r_{\rm nc}e^{-r_{\rm nc}T^{\rm c}}}{1+k_{\rm
  c}/r_{\rm c}+k_{\rm nc}/r_{\rm nc}}\frac{k_{\rm B}}{r_{\rm B}},\label{eq:B}
\end{align}
Assuming again that most bindings longer than $T^{\rm c}$ are cognate,
we solve Eqs.~(\ref{eq:A}, \ref{eq:B}) for the on-rates
\begin{align}
  k_{\rm c}&=\frac{\bar{B}e^{r_{\rm c}T^{\rm c}}r_{\rm c}r_{\rm
             B}}{k_{\rm B}}\left(1+\frac{\bar{A}}{k_{\rm A}/r_{\rm
             A}-\bar{A}}\right),\label{KPR_A1}\\
  k_{\rm nc}&=\left[\frac{\bar{A}}{k_{\rm A}/r_{\rm A}-\bar{A}} -\frac{\bar{B}e^{r_{\rm c}T^{\rm c}}r_{\rm
              B}}{k_{\rm B}}\left(1+\frac{\bar{A}}{k_{\rm A}/r_{\rm
              A}-\bar{A}}\right)\right]r_{\rm nc}.\label{KPR_A2}
\end{align}
The corrections of the form
${\bar{A}}/({k_{\rm A}/r_{\rm A}-\bar{A}})$ appear because bindings
only happen to unbound receptors, as emphasized in
Ref.~\cite{endres2009maximum}. However, these nonlinear relations are
still hard to implement with simple biochemical components. We solve
this by further assuming
$\epsilon= \bar{A}/(k_{\rm A}/r_{\rm A})\ll1$, which is true if the
receptor is mostly unbound (both on-rates are small compared to the
respective off-rates). This gives
\begin{align}
  k^{\rm KPR}_{\rm c}&\approx\frac{\bar{B}e^{r_{\rm c}T^{\rm c}}r_{\rm c}r_{\rm
                       B}}{k_{\rm B}},\label{KPR1}\\
  k^{\rm KPR}_{\rm nc}&\approx\left(\frac{r_{\rm A}\bar{A}}{k_{\rm A}} -\frac{\bar{B}e^{r_{\rm c}T^{\rm c}}r_{\rm
            B}}{k_{\rm B}}\right)r_{\rm nc}.
\label{KPR2}
\end{align}
These equations are analogous to Eqs.~(\ref{kca}, \ref{knca}).  They
are easy to realize biochemically (cf.~Fig.~\ref{fig:KPR}):
$k_{\rm c}$ is related to the concentration of the proofread species B
by a rescaling, and $k_{\rm nc}$ comes from subtracting rescaled
versions of A and B from each other. The subtraction can be done by
the third species C, activated by A and suppressed by B. Since
$\epsilon\ll 1$, then $\bar{A}$ and $\bar{B}$ are small, and many such
activation-suppression schemes are linearized as the subtraction
\cite{mugler15}.

The bias of $k^{\rm a}_{\rm c,nc}$ due to long, but noncognate binding
events, Eq.~(\ref{bias}), carries over to $k^{\rm KPR}_{\rm c,nc}$.
However, there is an additional contribution since the time to
traverse the intermediate states is random. Thus $T^{\rm c}$ has some
variance $\sigma^2_{T^{\rm c}}$ \cite{Bel10,Cheng:2013kx}. This
variability changes the rate of occurence of long biding events, but
they are still rare, nearly independent, and
Poisson-distributed. Denoting by $\langle\cdot\rangle$ the averaging
at a fixed $T^{\rm c}$, and by $\overline{\cdot}$ the averaging over
$T^{\rm c}$, we get
\begin{equation} \frac{\overline{\langle n_{\rm l}\rangle}}{k_{\rm c}}
  \approx\frac{n}{k_{\rm{c}}+k_{\rm{nc}}}e^{-r_{\rm{c}}\bar{T}^{\rm c}
    +\frac{1}{2} r_{\rm{c}}^2 \sigma_{T^{\rm c}}^2}.
\end{equation}
Thus $\sigma_{T^{\rm c}}^2 $ effectively renormalizes the cutoff to
$\bar{T}^{\rm c}-\frac{1}{2} r_{\rm{c}} \sigma_{T^{\rm c}}^2$, which
is independent of the on-rates.  Replacing $T^{\rm c}$ in
Eqs.~(\ref{KPR1}, \ref{KPR2}) by its renormalized value, which is an
easy change in the scaling factors, removes this additional bias due
to the random $T^{\rm c}$ in the KPR scheme.

Since long bindings are rare, the variance of the KPR estimator is
dominated again generally by $\bar{B}$, but not $\bar{A}$. The
intrinsic stochasticity in production of molecules of B contributes to
the variance. However, this contribution can be made arbitrarily small
by increasing $k_{\rm B}$, and we neglect it here.  A larger
contribution comes from the random number of long bound intervals and
a random duration of each of them.  To calculate this, in the limit of
rare long binding events, we use well-known results in the theory of
noise propagation in chemical networks \cite{Paulsson:2005uz}
\begin{multline} \frac{\sigma^2_{\rm
      B}}{\bar{B}^2}\approx\frac{\left(1+k_{\rm c}/r_{\rm
        c}+k_{\rm nc}/r_{\rm nc}\right)e^{r_{\rm c}T^{\rm
        c}-\frac{1}{2} r_{\rm{c}}^2 \sigma_{T^{\rm
          c}}^2}}{k_{\rm c}(1/r_{\rm c}+1/r_{\rm B})}\\
= \frac{e^{r_{\rm c}T^{\rm
        c}-\frac{1}{2} r_{\rm{c}}^2 \sigma_{T^{\rm
          c}}^2}}{k_{\rm c}(1/r_{\rm c}+1/r_{\rm B})}+O(\epsilon).
\label{KPR_var}
\end{multline}
This is a direct analog of Eq.~(\ref{eq:var_app}).

{\em Discussion.} The realization of
Refs.~\cite{endres2009maximum,siggia2013decisions} and others that the
detailed temporal sequence of binding and unbinding events carries
more information about the ligand concentration than the mean receptor
occupancy is a conceptual breakthrough.  It parallels the realization
in the computational neuroscience community that precise timing of
spikes carries more information about the stimulus than the mean
neural firing rate
\cite{Strong:1998vb,Reinagel:2000ts,Liu,Nemenman:2008ft,Fairhall:2012hk,Tang:2014bx},
and it has a potential to be equally impactful. This extra information
when measuring one ligand concentration with one receptor
\cite{endres2009maximum} amounted to increasing the sensing accuracy
by a constant prefactor, or, equivalently, getting only a finite
number of additional bits from even a very long measurement
\cite{Bialek:2001wv}. In contrast, here we show that two
concentrations can be measured with one receptor with the variance
that decreases inversely proportionally to the number of observations,
$n$, Eq.~(\ref{eq:var_app}), or to the integration time,
$1/r_{\rm B}$, Eq.~(\ref{KPR_var}), so that the accuracy is only a
(small) prefactor lower than would be possible with one receptor per
ligand species. Asymptotically, this doubles the information obtained
by the receptor \cite{Bialek:2001wv}.

In principle, one can measure more than two concentrations similarly,
as long as all species have sufficiently distinct off-rates. While the
error (the variance for the ML estimator, and both the bias and the
variance for the approximate and the KPR estimators) would grow with a
larger number of ligand species, this would still represent a dramatic
increase in the information gained by the receptor that keeps track of
its precise temporal dynamics, rather than just the average binding
state.

Crucially, such improvement would not be possible without the
cross-talk, or binding among noncognate ligands and receptors.
Normally, the cross-talk is considered a nuisance that must be
suppressed \cite{McClean:2007ef,Laub}. Instead we argue that
cross-talk can be beneficial by recruiting more receptor types to
measure concentration of the same ligand. In particular, this allows
having fewer receptor than ligand species, potentially illuminating
how cells function reliably in chemically complex environments with
few receptor types. Further, the cross-talk can increase the dynamic
range of the entire system: a ligand may saturate its cognate
receptor, preventing accurate measurement of its (high) concentration,
but it may be in the sensitive range of non-cognate receptors at the
same time. Finally, the increased bandwidth may lead to improvements
in sensing a time-dependent ligand concentration \cite
{Lalanne:2015ut,siggia2013decisions}. We will explore such
many-to-many sensory schemes, extending ideas of Ref.~\cite{morozov}
to tracking temporal sequences of activation of receptor and to
varying environments in forthcoming publications.

While the exact maximum likelihood inference of multiple
concentrations from a temporal binding-unbinding sequence is rather
complex, we showed that when the cognate and the non-cognate off-rates
are substantially different, there is a simpler, approximate, but
accurate inference procedure. In various immune system problems,
$r_{\rm nc}/r_{\rm c}\sim 5$, which would allow the approximation to
work.  Moreover, when the receptor is not saturated and spends most of
its time unbound, this inference can be performed by biochemical
motifs readily available to the cell. Namely, one needs two branches
of activation downstream of the receptor, with one of them having a
kinetic proofreading (KPR) time delay, and then an estimate of the
difference of activities of the branches. This suggests a possible
signal estimation role for the KPR scheme in addition to the more
traditional signal detection one
\cite{Francois:2013gx,Lalanne:2015ut,siggia2013decisions}. Such
branching and merging of signaling pathways downstreams of a receptor
is common in signaling \cite{Laub,Cheong:2011jp}. Thus exploring the
function of such complex organization in the context of estimation of
multiple signals with cross-talk is in order.

In summary, monitoring precise temporal sequences of receptor
activation/deactivation opens up new and exciting possibilities for
environment sensing by cells.

{\bf Acknowledgements:} This work was supported in part by the James
S.\ McDonnell foundation and the National Science Foundation.

\bibliography{receptors}

\end{document}